\documentclass{vldb}
\usepackage{usenix2019_v3}

\usepackage{graphicx}
\usepackage{hyperref}
\usepackage{balance}  
\usepackage{subcaption}
\usepackage{algorithm2e}
\usepackage{xspace}

\begin{document}

\newcommand{\pip}{\texttt{pip}\xspace} 
\newcommand{\todo}[1]{\textbf{TODO: #1}}
\newcommand{\parhead}[1]{\medskip \noindent \textbf{#1~~}}


\title{I Know What You Imported Last Summer: A study of security threats in the Python ecosystem}


\numberofauthors{1} 

\author{
%
%
\alignauthor
Aadesh M. Bagmar, Josiah Wedgwood, Dave Levin, Jim Purtilo\\
       \affaddr{University of Maryland}\\
       \affaddr{College Park, MD}\\
       \email{(aadesh, jwedgwo, dml, jim)@umd.edu}
}

\date{09 December 2019}
\maketitle

\begin{abstract}
The popularity of Python has risen rapidly over the past 15 years. It is a major language in some of the most exciting technologies today. This popularity has led to a large ecosystem of third party packages available via the pip package registry which hosts more than 200,000 packages. These third party packages can be reused by simply importing the package after installing using package managers like pip. The ease of reuse of third party software comes with security risks putting millions of users in danger. In this project, we study the ecosystem to analyze this threat. 

The mature ecosystem of Python has multiple weak spots that we highlight in our project. First, we demonstrate how trivial it is to exploit the Python ecosystem. Then, we systematically analyze dependencies amongst packages, maintainers and publicly reported security issues. Most attacks are possible only if users install malicious packages. We thus try to analyze and evaluate different methods used by attackers to force incorrect downloads. We quantify our ideas by estimating the potential threat that can be caused by exploiting a popular Python package. We also discuss methods used in the industry to defend against such attacks. 
\end{abstract}

\section{Introduction}

Python is one of the most popular languages and is being used in varied areas like machine learning, natural language processing, medical technologies and web development. Python is supported by a popular package manager, \textbf{pip}, which provides a platform to share and reuse code written by third party developers. Pip supports downloading packages from pypi.org, an online database of Python packages and installs the package by running a special script in the downloaded file called setup.py. The setup.py may instruct pip to resolve dependencies by recursively installing required packages. Since it's inception in 2005, PyPI has steadily grown to over 200,000 packages and is a primary source of widely used third-party packages. Thus, pip is an important part of the Python ecosystem. 

The PyPI ecosystem, being open by design allows arbitrary users to share and reuse code. Installing a third party package simply involves using pip which downloads the package and recursively downloads its dependencies. Using a package is trivial as well where a user needs to simply invoke a single line command. The ease of the infrastructure unsurprisingly comes with security risks. Recently, in December 2019, Python security team found two Python packages stealing SSH keys from users. In October 2018, malicious packages were found mining bitcoin on users' systems. In 2017, 10 rogue packages were found impersonating popular Python packages. Researchers have experimented \cite{PyPIBotnet}, \cite{TSquat} by uploading builtin Python packages (which users don't need to download since they are bundled with Python) on PyPI and have observed close to 500k downloads. These incidents show us the importance of analyzing the Python ecosystem.

Package managers and ecosystems have been explored in the past. e.g. \cite{zimmermann2019small} studies the NPM ecosystem. We do a similar extensive study for the Python ecosystem. Additionally, we highlight issues which allow an attacker to run malicious code at package installation time and at runtime, provide empirical data about methods used by attackers, quantify the impact of the defenses used in the industry and lastly identify potential license violations caused due to transitive dependencies. It's also important to note that Python is majorly used for back-end development and runs on the host system often without any sandboxing. The packages are also often installed using root permissions.

In this paper, we systematically study the security risks in the Python ecosystem by analyzing the triviality of reuse of packages, their dependencies, maintainers of packages and publicly reported security issues. Our study involves 206,296 packages, 1.5 million releases, 387,867 authors/maintainers and 600+ publicly known security issues in Python. About 34,000 packages statically indicate dependencies in the PyPI metadata. The overall finding is that the security loopholes in Python package management are severe and can be exploited. 

\parhead{Contributions}
Our paper makes the following contributions:


\begin{itemize}

\item \textbf{Privilege escalation at install time (\S\ref{sec:security_risk}):} We demonstrate a method which may allow an attacker to \emph{trivially} run arbitrary code with root privilege during the installation process.  We analyze the presence of such an attack on all our packages and find that 0.28\% packages execute a function, other than setup() and 0.78\% packages import other packages at \textbf{install} time.

\item \textbf{Trust metrics of PyPI (\S\ref{sec:trust_metrics}):} We analyze the amount of implicit trust and \emph{reach} of packages and maintainers in the PyPI ecosystem.  Much like with NPM~\cite{zimmermann2019small}, we find that an average Python package implicitly trusts 14 other packages while installation. We find the average number of packages a particular python package reaches has been steadily rising over the years and is \textbf{247} for the top 10,000 packages. Similarly, the average maintainer reach has been increasing over time as well, with maintainers reaching \textbf{338} packages as of 2019. 

\item \textbf{Package impersonation attacks (\S\ref{sec:impersonation}):} We provide what we believe to be the first analysis of \emph{impersonation attacks} on software packages.  We analyze the names of the packages based on previously known attacks. We also report on the efforts of a group of apparently benevolent developers who are proactively impersonating other popular packages as a means of \emph{defense}, as well as the first analysis of downloading typo-squatted packages.

\item \textbf{License violations (\S\ref{sec:license}):}  We also provide what we believe to be the first analysis of OSS license violations caused due to incompatible imports of packages. We report \textbf{672} potential violations caused due to import of packages with restrictive \textbf{GPLv3} license by packages with permissive license and \textbf{832} violations due to import of packages with \textbf{LGPLv3}.

\end{itemize}

Overall, to the best of our knowledge, there is no prior work which combines empirically studying the PyPI index, potential security risks in the ecosystem, impact of attacking certain packages/maintainers, license violations and defenses against such attacks. 

\parhead{Organization} The paper is organized as follows. Section \S\ref{sec:data} discusses the dataset and provides a brief analysis of it. Section \S\ref{sec:lifespan} explains the life cycle of a package in PyPI. Section \S\ref{sec:security_risk} explains our exploit and analyzes packages for its presence. The section also discusses the potential impact of attacking a package or hijacking a maintainer's account. Section \S\ref{sec:pypisecurity} explains the impact with an example. Section \S\ref{sec:impersonation} discusses package impersonation attacks and explains methods of defense. Ultimately, Section \S\ref{sec:license} discusses OSS license violations.

\begin{figure}
    \centering
    \includegraphics[width=\linewidth]{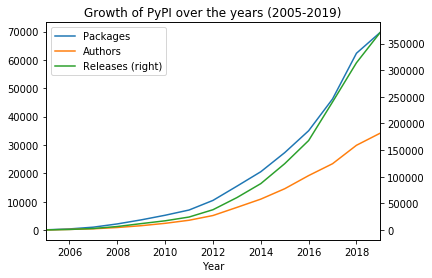}
    \caption{Growth of PyPI over the years between packages, releases, and authors.}
    \label{fig:pypi_growth}
\end{figure}

\section{Data Collection and Overview}
\label{sec:data}

The data for this study was collected by scraping the PyPI index, querying Google's BigTable and the PyPI stats website. Additionally, we also scraped the Safety DB free Database of vulnerabilities and the publicly available CVE dataset. We describe our collection methodology in detail below.

\subsection{Data Collection}
We collected the data from the following sources:

\parhead{PyPI packages} - PyPI.org is a public repository of Python packages where users can find, install and publish packages. It hosts more 200k packages with more than 1.5 million releases. Every package contains metadata associated with it. The metadata fields include information like \textit{Package Name, Author Email, Maintainer Email, Static Dependencies, License, Release data} which includes information like various versions and their release dates.

\parhead{Package download statistics} The Linehaul project \cite{linehaul} makes PyPI download statistics publicly available via Google BigQuery~\cite{bigquery}.\footnote{PyPIstats.org also makes these data available, but rate-limits queries.}  We queried this to fetch information regarding the downloads. Due to slight differences in names, we could match the metadata collected from PyPI.org to the information collected from BigQuery for \emph{206,296 packages}. Due to name mismatch in some packages we could match download information for \emph{148,644} packages. Only these packages were used while analyzing download counts and all packages were used in other cases.

\parhead{Vulnerability datasets}
Safety DB \cite{SafetyDB} provides a mapping between commonly identified vulnerabilities and the Python packages having those vulnerabilities. The dataset directly also gives information about the type of attack, which releases were affected and the date and time when the vulnerability was identified. Safety DB contains information for \emph{617 packages}.  We also collected and extracted information from the publicly available CVE database.

\subsection{Data Overview}
Once our entire dataset was pushed into the Neo4j graph, we queried the graph to extract information about PyPI. Here is a brief analysis of the ecosystem.

\begin{figure*}
    \begin{subfigure}{.33\textwidth}
        \centering
        \includegraphics[width=\linewidth]{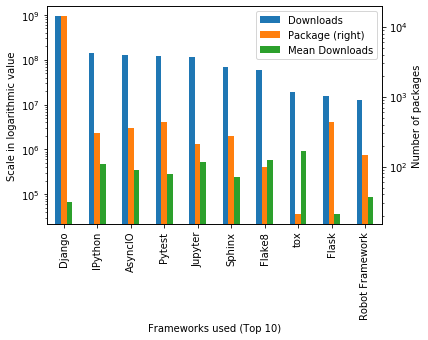}
        \caption{Top frameworks for which packages are built}
        \label{fig:Framework}
    \end{subfigure}
    \begin{subfigure}{.33\textwidth}
        \centering
        \includegraphics[width=\linewidth]{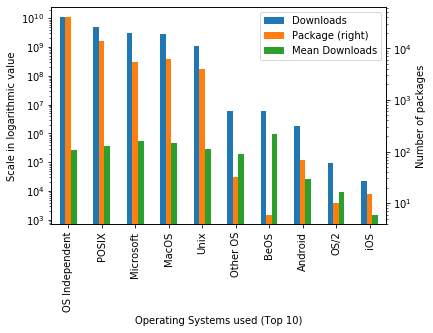}
        \caption{Categorizing by supported operating systems}
        \label{fig:OS}
    \end{subfigure}
    \begin{subfigure}{.33\textwidth}
        \centering
        \includegraphics[width=\linewidth]{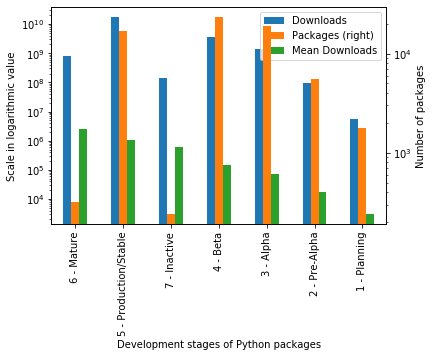}
        \caption{Categorizing by Package development status}
        \label{fig:DevStat}
    \end{subfigure}
    \caption{Classifying the packages based on frameworks, operating systems and development status. All the graphs have scales on both left and right. The right scale is used to portray download counts while the scale on the left is used for package downloads and average downloads.}
    \label{fig:PyPI classification}
\end{figure*}

\subsubsection{Growth of PyPI}
PyPI has sustained a double digit growth rate for the last 13 years. On average, PyPI has seen more than 51\% releases per year and 31\% new authors have increased per year.
Figure \ref{fig:pypi_growth} shows the growth trend for number of packages, authors and releases over the years. We can see that the growth of number of packages is almost twice as that of authors. 

Currently, PyPI has \textbf{206,296 projects}, \textbf{1,554,933 releases}, \textbf{387,867 users} and total downloads amounting to over 100 million. Our directed graph contains information about \textbf{198,202 packages} and has \textbf{230,566 edges} showing static dependencies. 

\subsubsection{Analysis of the ecosystem}
Based on the metadata collected, we classify the ecosystem by various categories. We can use this to understand the attack vector better. This helps us identify the likely targets in case the ecosystem is attacked.

We analyze the packages as follows: 

\parhead{Framework Types} Django is the most popular (57\%) framework for which packages are built in PyPI. Figure \ref{fig:Framework} shows the trend. The most number of packages and most number of downloads both indicate Django's massive popularity. This also explains why attacks have been launched on Django before.

\parhead{Topics} Topics indicate purpose of building a package. As expected, 37\% packages are built for Software Development. The graph also shows that 1.6\% packages are used for security applications which makes this study extremely important. 

\parhead{Supported Operating Systems} - Most packages (58\%) are OS independent in PyPI and can run cross platform. However, there are 10\% MacOS and Microsoft specific packages. Figure \ref{fig:OS} shows the trend in PyPI. The open nature of Python makes it possible for packages to run independent of the operating system makes it a bigger threat.

\parhead{Development Status} A majority of the packages are in either in Alpha (34\%) or Beta (28\%) release. Less than 25\% packages call themselves Production ready. Figure \ref{fig:DevStat} shows the trend. However, Production / Stable packages have the most number of downloads cumulatively. Mature packages have the highest average number of downloads. 

\parhead{Intended Audience} - Most packages (66\%) cater to Developers. 12\% are built for scientific research.

\section{Life Span of a package on PyPI}
\label{sec:lifespan}
Packages on PyPI go through three developing stages before being installed onto a given host.

\parhead{Development} Python packages are specially structured directories containing a script ``setup.py", the LICENSE file, a README for the package, and more sub-directories for each package module. The script setup.py contains the package metadata. Each module is marked by the script "\_\_init\_\_.py", which can simply be an empty file for pip.

\parhead{Distribution} Packages are uploaded to PyPI via a two-step process. They start by building a source distribution by executing setup.py with the "sdist" or "bdist" arguments. Once the packages are built, they are uploaded via Twine to the registry. Distributions are versioned archived files containing Python packages. Making either a source distribution or a build distribution leads to the same upload procedure using Twine.

\parhead{Maintenance} When updating packages published on PyPI, each update is uploaded as a new release. PyPI only requires the version number to be incremented upon every new package release.

Packages live on and are freely available for download on PyPI until either the author, maintainer, or a PyPI developer removes the project. Packages that are deleted from PyPI cannot be redistributed on the platform under the same name, so removing a package also removes the package name from PyPI. However, the deleted packages may still live on the \textbf{Python.filehosted.org} server. 

\section{Security risk in the ecosystem}
\label{sec:security_risk}
In this section we discuss a simple method to exploit pip's setup.py architecture and the dangers it entails. 

\subsection{Executing arbitrary code}
While setup.py and \_\_init\_\_.py serve specific functions for pip and the package development process, both scripts can contain arbitrary Python code. Each script is executed by the Python interpreter at varying points in development:

\begin{enumerate}
    \item \textbf{setup.py} - Executed upon package installation, and may be executed multiple times during package installation. The user may run it using \textbf{sudo} command too.
    
    \item \textbf{\_\_init\_\_.py} - Executed whenever a package is imported.
\end{enumerate}

The exploit works when installing a source distribution zip or tarball, or while installing in editable mode from a source tree. This is likely to happen if users manually download Python packages from github repositories and try to install it from the source try by executing \textbf{Python setup.py install}. Additionally, the attack can also be executed if a package simply imports the malicious package. The fundamental architecture seems to be broken since it is extremely easy to execute malicious code. This is not true in other languages like C, C++, Java where users need to call functions after importing header files to execute code present in the third party libraries. We added the following code to setup.py which enables us to get complete remote access of a users system.

\parhead{Exploit} Since some Linux distributions come pre-installed with Python, some users may run installation as \textbf{sudo}. Adversaries may not need to force privilege escalation! We \emph{modified the setup.py} file for a package to remotely connect to a listening server and establishes a 2 way connection. The code takes input from the server and pipes it to the \textbf{/bin/bash} of the system where the package is running. Thus, a remote user has complete access of the system. The relevant code for doing so is shown below. The code snippet has been instructed to run \textbf{Post Installation.} It runs as a \emph{separate process}. Thus, a user would \textbf{not} notice any difference. The package will work as expected to make it even more difficult to spot. The attacker has \emph{full access} and can even turn on a users' microphone/camera. The algorithm is mentioned below:

\begin{verbatim}
    class PostInstallCommand(install):
      """Post-installation for installation 
      mode."""
    
      def run(self):
          # Create a socket connection
          s = socket.socket(socket.AF_INET,
          socket.SOCK_STREAM)
          s.connect(("<attacker-IP>", 
          <attacker-port>))
          
          # Create multiple file descriptors
          os.dup2(s.fileno(), 0)
          os.dup2(s.fileno(), 1)
          os.dup2(s.fileno(), 2)
          
          # Stream input to /bin/bash
          p = subprocess.Popen(["/bin/bash", "-i"],
          close_fds=True)
\end{verbatim}

\subsection{Analyzing setup.py files across packages}

\begin{figure}
    \centering
    \includegraphics[width=\linewidth]{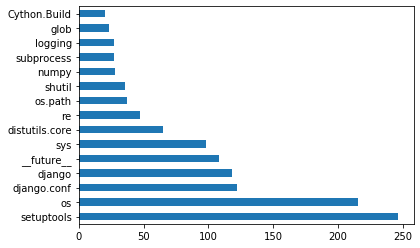}
    \caption{Top 15 most imported packages at install time. Note that packages like os, sys, shutil, feature, glob in the list.}
    \label{fig:Installtime}
\end{figure}

It is likely that users may decide to clone the GitHub project and run \emph{Python setup.py install}. This script can run arbitrary code on the user's machine. Thus, it is important to analyze setup.py files across the ecosystem. For this experiment, we scraped pypi to get the latest wheel files for every package. We found that 51\% packages used \emph{setup.py} for installation. We then downloaded the \emph{wheel} files for their latest release and extracted setup.py files from all those packages. Out of them, 642 (\textbf{0.39\%}) packages import other packages at install time. Figure \ref{fig:Installtime} shows the most imported packages. The graph shows that packages like os, shutil, subprocess, glob are used frequently which are capable of executing system calls and performing directory traversals. We also found that 220 (\textbf{0.28\%}) packages executed functions other than \emph{setup()}, the standard function required by setup.py. \textbf{99\%} of the setup.py files have a simple setup() function suggesting that we could replace the executable file with a simple manifest. This manifest could be taken as an input by some other program which could run static tests before executing it. 

Fortunately, none of the packages we surveyed had any malicious code running at install time or were using \emph{PostInstall} commands.

\subsection{Impact of exploiting PyPI}
\label{sec:trust_metrics}
In this section we discuss four main factors that help us understand the impact of exploiting a Python package or compromising a maintainer account. Our findings are along the same lines as those observed by \cite{zimmermann2019small} for the npm ecosystem. However, we also discuss the parameters below in terms of real-world vulnerabilities.

\subsubsection{Package Reach}
In Python any package can import any other third party package by listing it as a dependency. A dependency maybe direct or transitive. A direct dependency indicates that a package x directly imports another package y. Packages x and y share a direct dependency in this case. Packages may also import packages transitively where a package x imports another package which in turn imports y. Packages X and Y are transitively dependent in this case. Transitive dependencies may go on for multiple hops. 

We define the \textit{Package Reach} for a given package to be the number of other packages that explicitly require it either transitively or directly. Packages with a higher Package Reach demonstrate a larger attack vector should they become malicious. Such packages can create a ripple effect and spread vulnerabilities to a much larger set of packages. This term has been previously described and used in \cite{zimmermann2019small}.

\begin{figure}
    \centering
    \includegraphics[width=0.5\textwidth]{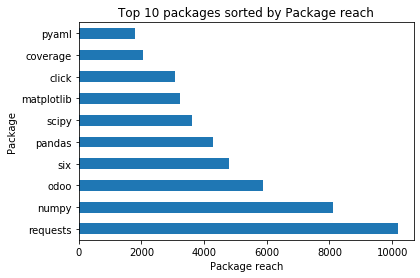}
    \caption{Top 10 packages sorted by their total package reach.}
    \label{fig:package_reach}
\end{figure}

Figure \ref{fig:package_reach} shows the top-ten packages with the highest Package Reach. Each of the packages listed are prime targets for adversaries to spoof or hijack because of their tremendous reach across the ecosystem. If any dependent packages also  do not specify any version number, then a single new release uploaded to PyPI can be installed by any user who happens to update their dependencies.

\begin{figure}
    \centering
    \includegraphics[width=\linewidth]{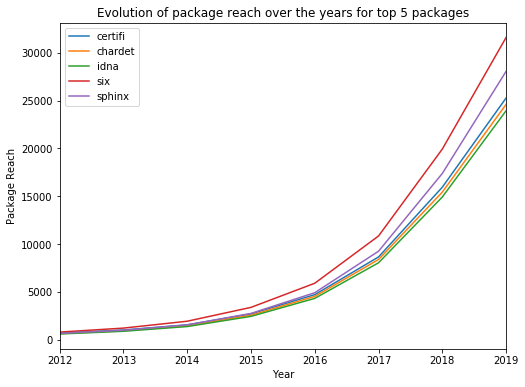}
    \caption{Evolution of package reach over the years for top 5 packages.}
    \label{fig:top5pr}
\end{figure}

\begin{figure}
    \centering
    \includegraphics[width=\linewidth]{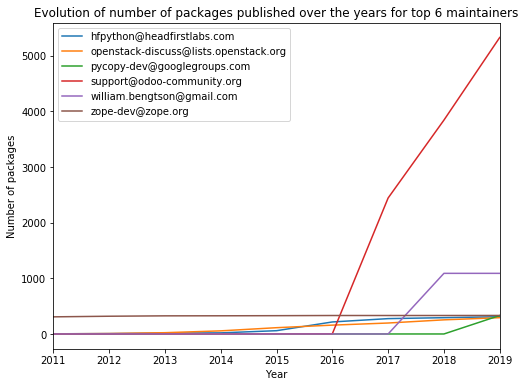}
    \caption{Evolution of Top 6 maintainers sorted by number of packages they own over the years. Note the sudden start to odoo community in 2016.}
    \label{fig:top_5_maintainers}
\end{figure}

\begin{figure}
    \centering
    \includegraphics[width=\linewidth]{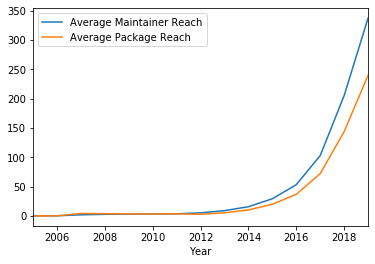}
    \caption{Evolution of Package Reach and Maintainer Reach for the top 10,000 packages on PyPI over time.}
    \label{fig:pr_mr_overtime}
\end{figure}

The package reach has increased at a non-linear rate over the years. Some highly popular packages reach more than 30,000 other packages, making them a prime target for attacks. This problem as seen in Figure \ref{fig:pr_mr_overtime} has been aggravating over the past few years. The top packages on PyPI rarely even included more than five packages before 2013.Every year thereafter we can observe an exponential growth in the package reach of these popular packages. 

The top 5 packages show a similar rise in package reach around 2016-2017. The trend can be seen in Figure \ref{fig:top5pr}. These packages reach between 28,000 and 32,000 other packages for transitive dependencies upto depth 5 making them highly attractive targets. Drawing similar conclusions as drawn by \cite{zimmermann2019small} about growing package reach in NPM, we can say that an increase in the average package reach implies an increase in the attractiveness of attacks that rely on dependencies.

\subsubsection{Maintainer Reach}
Like Package Reach, we also use the \textit{Maintainer Reach} definition from \cite{zimmermann2019small}. The top-ten maintainers with the highest Maintainer Reach is shown in Table \ref{tab:maintainer_blast}.

\begin{table}
\centering
    \begin{tabular}{r|l}
    \textbf{Package Count} & \textbf{Maintainer Email} \\ \hline
    5331 & support@odoo-community.org  \\
    1090 & william.bengtson@gmail.com \\
    347 & hfPython@headfirstlabs.com \\
    334 & zope-dev@zope.org \\
    332 & pycopy-dev@googlegroups.com   \\
    293 & openstack-discuss@lists.openstack.org \\
    205 & micro-Python@googlegroups.com   \\
    186 & openstack-dev@lists.openstack.org   \\
    181 & aliyun-developers@list.alibaba-inc.com  \\
    167 & plone-developers@lists.sourceforge.net \\
    \hline
    \end{tabular}
    \caption{Maintainer emails with the most number of packages in direct control. Note the non-corporate/community email address from William Bengston.}
    \label{tab:maintainer_blast}
\end{table}

These maintainer accounts are also high value targets for adversaries because gaining access to one of these accounts offers a large attack surface on packages under the account's ownership. As we observed in Figure \ref{fig:pypi_growth}, the number of authors/maintainers grow at a slower rate than the number of packages. This implies that an author may be maintaining one or more packages. If an attacker manages to get the credentials to an authors account, he could attack multiple packages. 

Figure \ref{fig:top_5_maintainers} shows number of packages published over the years by top 6 maintainers. Most accounts probably refer to groups of maintainers who use the same account. William Bengtson's packages are efforts to protect the community from typo-squatting attacks. His packages typo-squat popular packages and warn users if they try to download it. We discuss his efforts further in \ref{sec:Guardian}.

We can observe that the average maintainer reach on PyPI in Figure \ref{fig:pr_mr_overtime} follows a similar but larger trend to the package reach.

\begin{figure}
    \begin{subfigure}{.5\textwidth}
        \centering
        \includegraphics[width=\linewidth]{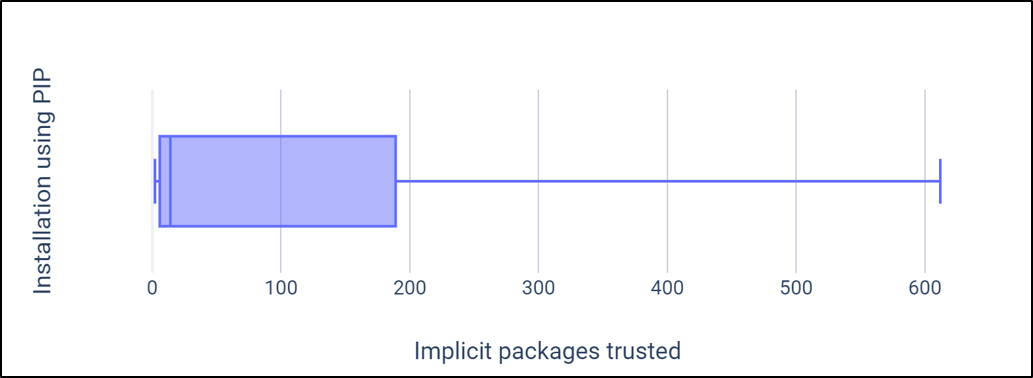}
        \caption{Quantiles for Implicitly Trusted Packages}
        \label{fig:ITP}
    \end{subfigure}
    \begin{subfigure}{.5\textwidth}
        \centering
        \includegraphics[width=\linewidth]{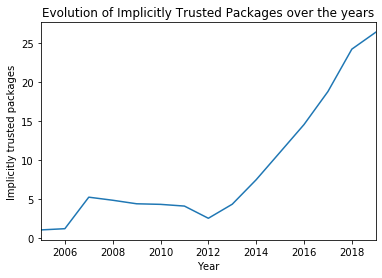}
        \caption{Evolution of implicitly trusted packages over the years.}
        \label{fig:evolution_itp}
    \end{subfigure}
    \caption{Rise and fall of implicitly trusted packages over time.}
\end{figure}

\subsubsection{Implicitly Trusted Packages}
We look at implicitly trusted packages (ITP), as defined in \cite{zimmermann2019small}. Since \pip (like npm) will also resolve transitive dependencies during installation time, we must also consider packages dependencies that are not explicitly stated in the package metadata. Our best approximation for measuring transitive dependencies in PyPI is to count the number of distinct nodes traversed while searching for the longest path from a given starting node. 

In our attempt to quantify the implicit package trust, we calculated that 50 percent of packages transitively require at least 14 other packages to be installed. In simple words, when a user installs one package, he/she is implicitly trusting 14 other packages on average. Figure \ref{fig:ITP} shows a box plot representation of implicitly trusted packages. In 2019, the average number of implicitly trusted packages go up to 27. 

Over the years, number of implicitly trusted packages have risen due to heavy code reuse. Figure \ref{fig:evolution_itp}, shows this growing trend for packages importing at least 1 other package. Since, the average number of exploitable bugs increase with every line of code, a constant increase in the number of implicitly trusted packages over the years, shows a chilling trend.

\subsubsection{Implicitly Trusted Maintainers}
\begin{figure}
    \centering
    \includegraphics[width=0.5\textwidth]{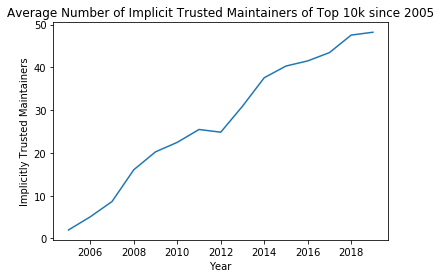}
    \caption{Evolution of implicitly trusted maintainers over time.}
    \label{fig:itm_overtime}
\end{figure}

We look at implicitly trusted maintainers (ITM), as defined in \cite{zimmermann2019small} for npm. We observe how the average ITM changes for the top 10,000 packages in PyPI overtime in Figure \ref{fig:itm_overtime}.

While Figure \ref{fig:pypi_growth} demonstrates that the number of packages outpaces the number of maintainers every year, Figure \ref{fig:itm_overtime} shows that maintainers have a steeply increasing influence on each other. The trend demonstrates the frequency of code reuse among packages since the very beginning of the ecosystem, which directly translates into how many attack vectors packages expose themselves to because of these dependencies.

The ITM is a measurement of how vulnerable a package is by other maintainer accounts. Compromising any maintainer account that is a part of a package's ITM set is also a compromise of that package's security. The compromised maintainer would be able to upload malware disguised as an update to their own packages, and other packages in the MR of the compromised account may update their dependencies to now include that malware. As described in \cite{zimmermann2019small}, any value higher than 20 for the ITM is a critical security concern for the package, As of 2019, the top 10,000 packages depends on code published by \textbf{49} maintainers.
\section{Security and PyPI}
\label{sec:pypisecurity}

\begin{figure}
    \centering
    \includegraphics[width=\linewidth]{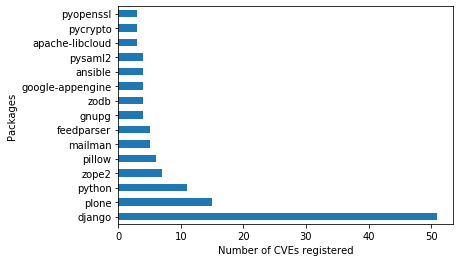}
    \caption{Top 15 packages with most CVEs listed against them in SafetyDB}
    \label{fig:CVE}
\end{figure}

Safety DB \cite{SafetyDB} provides a curated database of insecure Python packages, their releases and also lists CVEs associated with them. In our dataset, we have information about \textbf{648} packages and their \textbf{961} releases. There are \textbf{269} CVEs linked to \textbf{132 unique} packages. \hyperref[fig:CVE]{\ref{fig:CVE}} shows top 15 packages with most CVEs listed against them.

In our study, we evaluate the listed CVEs to get their publish date, description, severity score and last modification date. We assume the last modified date potentially indicates the day the vulnerability was fixed.

\begin{figure}
    \centering
    \includegraphics[width=\linewidth]{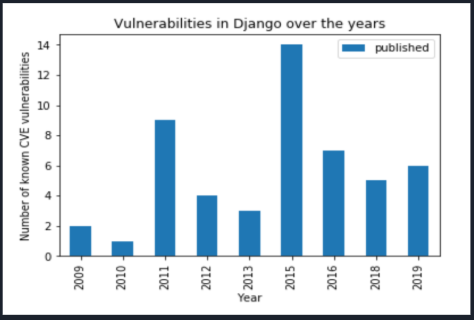}
    \caption{Number of known vulnerabilities in Django per year}
    \label{fig:YearsDjango}
\end{figure}

It is important to understand the cycle of vulnerability detection and fixing for popular packages. This can help developers identify security aware packages and understand the frequency at which packages should be updated. The time taken to fix a vulnerability indicates the window of opportunity that attackers have. Vulnerability timeline for a package affects all packages in its package reach. 
Our data enables us understand how vulnerabilities propagate over time. Here, we take an example of Django, a popular package and framework to understand this phenomenon. We also analyze who would be affected the most when a new vulnerability hits a popular package like Django. 

\subsection{Vulnerability propagation in Django}
\label{sec:DjangoVulnerabilityProp}
\begin{figure}
    \centering
    \includegraphics[width=\linewidth]{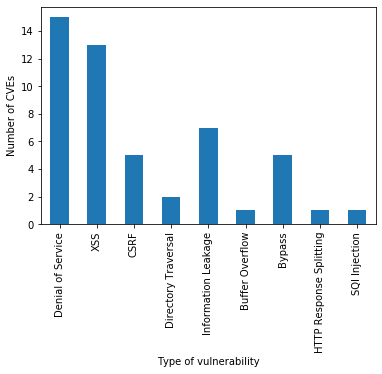}
    \caption{Types of vulnerabilities in Django}
    \label{fig:TypeDjango}
\end{figure}

Vulnerabilities once identified take a substantial amount of time to get fixed. That leaves a \textbf{window} open for an attacker to exploit the package. Most users pin their package versions. Thus, even if a library maintainer fixes their package, a user needs to explicitly update the version. In the following example, we discuss how vulnerability propagates in the ecosystem.

We chose Django because the package has more than 2 million downloads making it an ideal candidate to attack. It's package Reach is \textbf{15k} packages which can amplify the attack. Django was hit with a remote access vulnerability in its \textbf{1.8.9} release. Interestingly, more than 18\% of the reported vulnerabilities in Safety DB in \emph{other packages} are because users use an insecure version of Django.

\begin{figure}
    \centering
    \includegraphics{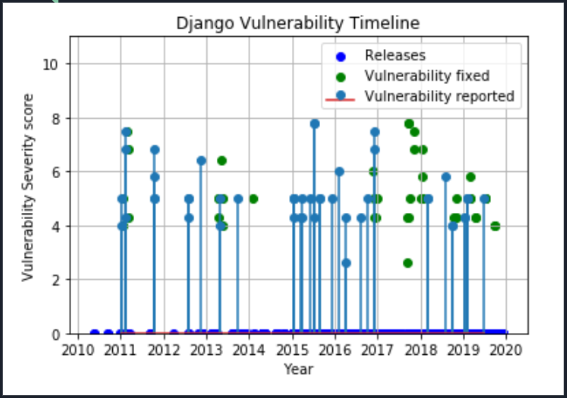}
    \caption{Timeline of spread of vulnerabilities in Django}
    \label{fig:django}
\end{figure}

Figure \ref{fig:django} shows releases, time at which a vulnerability was first reported and when was the vulnerability fixed. The y-axis demonstrates the severity of these vulnerabilities. Some vulnerabilities had an attack window of more than 3 years in Django.

Django over the years has had \textbf{355} releases and \textbf{51} CVEs have been reported with an average CVE severity score of \textbf{5.15}. Django has a massive package reach of 15,000 packages. The most common vulnerability that has been identified in Django has been \textbf{denial of service} attacks. Interestingly, \textbf{48} vulnerabilities have been identified in Django after version 1.8.
Figure \ref{fig:TypeDjango} shows types of vulnerabilities identified in Django. Denial of Service and XSS have been the most common attacks. \hyperref[fig:YearsDjango]{\ref{fig:YearsDjango}} shows a distribution of number of identified attacks in Django. The number of attacks are the highest in 2015 when several DoS vulnerabilities were found in the Django library. 

\begin{figure}
    \centering
    \includegraphics[width=\linewidth]{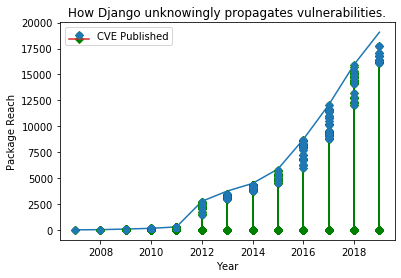}
    \caption{Rise of number of packages susceptible to getting infected by importing Django. The dots denote the number of vulnerabilities identified that year while the line shows the growing Package Reach of Django.}
    \label{fig:DjangoVuln}
\end{figure}

Figure \ref{fig:DjangoVuln} shows how packages Django unknowingly propagates vulnerabilities in packages that import Django. The graph plots Django's package reach over the years and also marks the dates when CVEs were published in Django. Since Django's package reach is growing exponentially, a new vulnerability in Django makes a large number of packages susceptible. Assuming that most packages pin versions, we studied what's the shortest time in which a package importing Django releases a new version of the package after a CVE in Django was fixed. This obviously gives a lower bound since the package may not be updating Django in it's release. The number would clearly be higher. 

\textbf{For the 64 packages that we tested, the average time difference between release of a vulnerability and release of a new version of a package importing a vulnerable package is \textbf{397 days}.}

\begin{figure}
    \centering
    \includegraphics[width=0.5\textwidth]{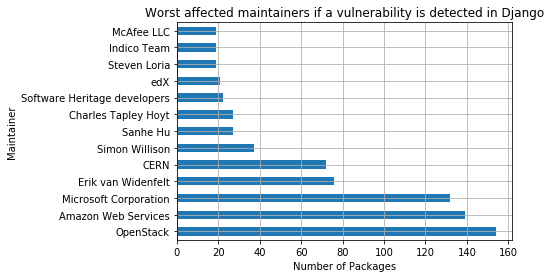}
    \caption{Worst affected maintainers, if a vulnerability hits Django}
    \label{fig:DjangoAuth}
\end{figure}

The impact factor of a vulnerability in Django is higher than it appears. A vulnerability in Django makes \textbf{154 OpenStack}, \textbf{139 AWS}, \textbf{132 Microsoft packages} and \textbf{72 CERN} susceptible to attack amongst many other packages. Interestingly, it also affects \textbf{19} packages belonging to \textbf{McAfee}, a company which provides security solutions. \hyperref[fig:DjangoAuth]{\ref{fig:DjangoAuth}} shows the top 13 worst affected maintainers who own packages importing Django.

Now that we can estimate the impact of an attack, we try to analyze the ways used by attackers to force malicious package downloads. We discuss some methods in the next section.

\section{Package Impersonation Attacks}
\label{sec:impersonation}
User centric attacks have been launched in PyPI in the past. They are mostly aimed at tricking the user to download a malicious package. Attackers have been known to use a combinations of the following attacks to fool users. Some of these methods are:

\parhead{TypoSquating} Relying on users making a spelling mistake while installing the package. e.g. numpy and numpi. A malicious package name needs to be similar enough to a legitimate package in order to create confusion for potential users, so we limited our search to package name pairs with a Levenshtein edit distance of at most three. We found that \textbf{41\%} packages have at least one other package within an edit distance of three. Within our dataset we observed \textbf{27,622} occurrences of package names with edit distance one, \textbf{180,928} occurrences of package names with edit distance two, and finally \textbf{849,972} occurrences with edit distance 3. 

This attack is widely known and contained by the pip community. Publications like \cite{typosquatI}, where 11 malicious packages were discovered and removed,  demonstrate that typosquating in this ecosystem is being actively monitored, with past attempts being removed from PyPI. We further analyze some recent typosquat attempts in section \ref{subsection:typosquat_exploits}.

\parhead{Altering Word Order} Changing the order in which the package names itself is another common way e.g. test-vision-client and client-vision-test. We found \textbf{277} pairs and \textbf{3} such triplets where packages used the same words but changed their order around a hyphenation or an underscore. \textbf{15\%} of these package pairs were owned by the same author which shows the effort put in by their authors to alleviate the confusion. Other packages (85\%) could be typo-squatting attempts or may create confusion for users who may end up downloading incorrect packages. These packages have a combined Package reach of \textbf{6078}.

\parhead{Python 3 vs Python 2} Another common attack is launching a packages by adding a number "3" to them. It confounds users by making them believe that 3 indicates support for Python3. e.g. Python3-dateutil. Around \textbf{1703} packages were found where packages were similar if \textbf{Python3} was removed from their names. About \textbf{1.6\%} packages have Python-3 associated with them.

\parhead{Removing Hyphenation} Removing hyphens from the original package to attack is also an interesting way. e.g. aws-cli and awscli. We found that \textbf{2296} packages had similar names if hyphenation was not considered. 

\parhead{Built in packages} Attackers have explored uploading default Python packages on Pypi. We found \textbf{32} packages which shared names with a built-in Python package. Out of these \textbf{7} packages were backports and were being maintained for people who haven't migrated to newer versions of Python.

\parhead{Joke Packages} We also noticed some Joke / Suspicious packages. e.g. urllib5, request. A user uploaded urllib5 mocking multiple libraries released by the same organization with the names urllib, urllib2, urllib3 and so on.

\begin{table}
    \begin{tabular}{l|l|l}
         \textbf{Malicious Package} & \textbf{Original Package} & \textbf{TypoSquat Method}\\ \hline
         acqusition & acquisition & Edit Distance \\
         crypt & crypto & Edit Distance \\
         setup-tools & setuptools & No Hyphenation \\
         urllib & urllib3 & Python3 attack \\
         djanga/ djago/ dajngo & django & Edit Distance \\
         easyinstall & easy\_install & No Hyphenation \\
         python-sqlite & squlite & Adding Python \\ 
         pkgutil & - & Standard Library \\ 
         subprocess & - & Standard Library  \\
         shutil & - & Standard Library \\ 
        \hline
    \end{tabular}
    \caption{Typosquatting attacks in the wild. Aggregates data collected from \cite{pytosquatting}. }
    \label{tab:pyto_squat}
\end{table}

Table \ref{tab:pyto_squat} shows a few real-world attacks. In the next section, we analyze a recent exploit which combines a couple of methods discussed above. We also show methods used by various companies which use these attack methods as a defense to protect their packages.

\subsection{Exploits in the wild - the Jellyfish attack}
\label{subsection:typosquat_exploits}

We study an interesting real-world exploit and try to analyze it. The Jellyfish package in Python recently faced a typosquating attack \cite{JellYFish}. Jellyfish is a package for phonetic string matching and has more than 300 thousand downloads. The typo-squatted package \textbf{JelIyfish} has more than 500 downloads which increased significantly when the same attacker uploaded another package named \textbf{Python3-dateutil} which internally imported the malicious JelIfish package. 

\begin{figure}
    \centering
    \includegraphics[width=\linewidth]{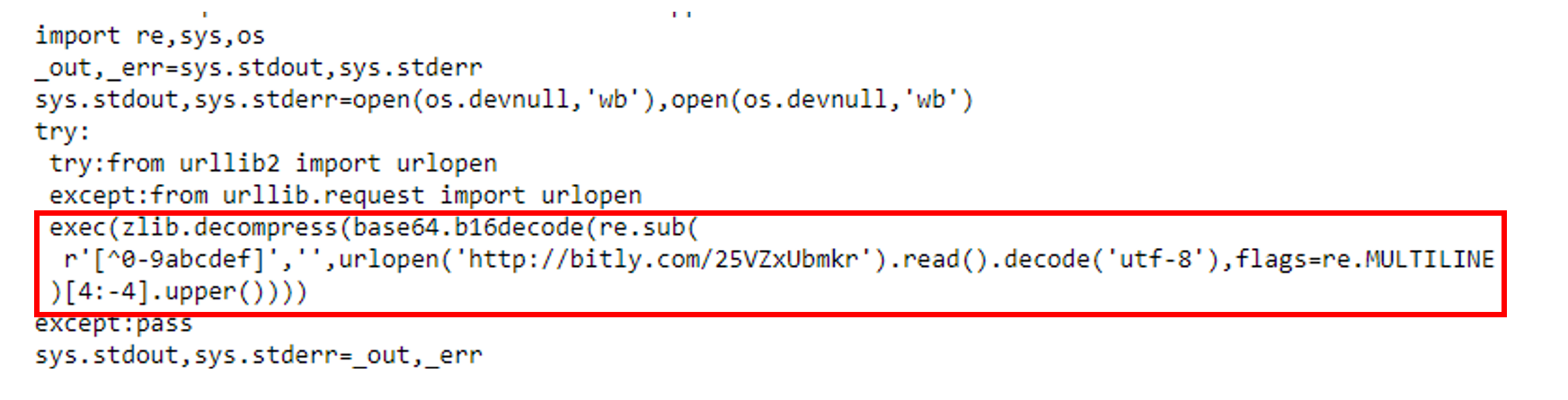}
    \caption{Decompressed malicious code from the JeIlyfish repository.}
    \label{fig:Decomp}
\end{figure}

The package \textbf{stole users' SSH keys} and streamed it to the server. The malicious code ran whenever a user \textbf{imported} the package. The code was present in its main library in compressed form which ran \textbf{exec} whenever the code is imported. On decompressing, the code shows that it downloads a binary from a URL hidden behind bitly.com which steals keys. The decompressed code can be found in \hyperref[fig:Decomp]{\ref{fig:Decomp}}

The original Jellyfish package had a Package Reach of just \textbf{17} which is why the package survived for over a year. However, once the attacker moved to attacking Python-dateutil which has a Package Reach of \textbf{21527}, the number of downloads rose exponentially. Once the attack came to light, the attacker deleted the repository where the malicious binary was hosted. Packages implicitly trust jellyfish while packages implicitly trust the Python-dateutil package.

We then focused our work to identify how frequently do users actually fall for such packages. While searching for packages, we constantly kept coming across projects belonging to an author (or a group of authors) who called themselves ``The Guardians". We describe the project in the next section. The project helps us compute a lower bound on the number of errors humans make while installing packages. 

\subsection{Using attacks as a defense - Prevention in the wild}
One simple method is to register multiple packages for one package where each package name enumerates and books common typosquating attempts. E.g. for a package with the name \textbf{Python-vagrant}, upload the same package with the name \textbf{Pythonvagrant} to prevent attackers from typo-squatting by removing the hyphenation. These methods have been used by various organizations in the past. The package owners typosquat their own packages to prevent attackers from typosquating them. We discuss some implementations below.

\subsubsection{The Guardian Project}
\label{sec:Guardian}
Our methods identified 1083 packages belonging to \textbf{The Guardians}. They created packages to protect people from getting exploited. Installing a Guardian package simply raises an error saying, ``Did you mean to install $<$legit-package$>$ instead?"

\begin{figure}
    \centering
    \includegraphics[width=\linewidth]{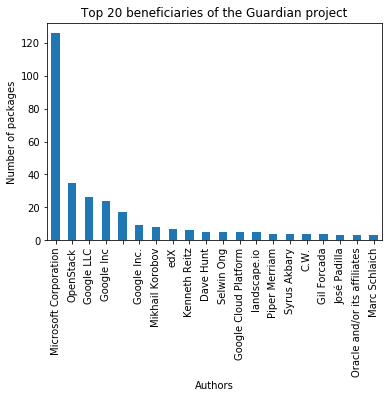}
    \caption{Top 20 authors who have benefited from the Guardian project}
    \label{fig:GuardianBeneficiaries}
\end{figure}

The Guardian project helps us evaluate a lower bound on how helpful such methods are. The Guardian project packages have more than \textbf{250k} downloads cumulatively. For a few packages, there are more than \textbf{46\%} hits to the typo-squatted package compared to the original package. Table \ref{tab:Guardians} shows top 15 packages which have been protected by the Guardians.

\begin{table*}
\footnotesize
    \centering
    \begin{tabular}{@{}r@{~~}|@{~~}l|r|r@{~~}r@{}}
     \multicolumn{2}{c|}{\textbf{Package Name}  } & \multicolumn{3}{c}{\textbf{\# Downloads}} \\
Original &  Typo-squatted &  Original &  Typo & (\% Total)  \\\hline
prompt-tool-kit &            prompttoolkit &    170 &            149 & (46.71\%) \\
trisdb-py &                      trisdbpy &     46 &    23 & (33.33\%) \\
trailblazer-aws &          trailblazeraws &     70 &    22 & (23.91\%) \\
django-simplecaptcha & djangosimplecaptcha &    171 &   29 & (14.50\%) \\
django-healthcheck &     djangohealthcheck &    126 &   21 & (14.29\%) \\
django-useragents &       djangouseragents &    292 &   29 & (9.03\%) \\
kms-vault &                       kmsvault &    282 &   20 & (6.62\%) \\
simple-crypt &                    simplecrypt &    168,031 & 6934 & (3.96\%) \\
pyqt5-tools &                    pyqt5tools &  245,395 &   8963 & (3.52\%) \\
django-daterangefilter    &               djangodaterangefilter &    1532 & 27 & (1.73\%) \\
scapy-Python3 &                 scapyPython3 & 92,257 & 826 & (0.89\%) \\
flake8-chart &                  flake8chart & 3811 & 31 & (0.81\%) \\
browsermob-proxy &            browsermobproxy &    315,336 & 2457 & (0.77\%) \\
ll-xist &                            llxist & 6120 & 32 & (0.52\%) \\
py-dateutil &                      pydateutil &    284,109 & 1173 & (0.41\%) \\\hline
\end{tabular}
    \caption{Top 15 packages saved by The Guardian Project. The percentage shows how many times the typo-squatted package was accessed}
    \label{tab:Guardians}
\end{table*}

The Guardians mainly protect against hyphenation attacks. \hyperref[fig:GuardianBeneficiaries]{\ref{fig:GuardianBeneficiaries}} shows that Microsoft, OpenStack and Google have the most number of packages protected by the Guardians. 

\begin{figure}
    \centering
    \includegraphics[width=\linewidth]{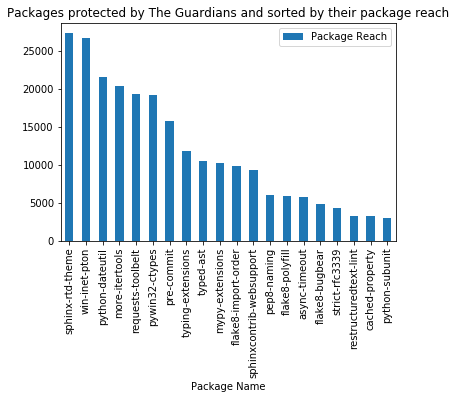}
    \caption{Top 20 packages protected by the Guardians ordered by their reach}
    \label{fig:ProtectionReach}
\end{figure}

\textbf{Package Reach} \\ The packages protected by the Guardians have a high package reach. \hyperref[fig:ProtectionReach]{\ref{fig:ProtectionReach}} shows the top 20 packages ordered by reach and protected by the Guardian project. These packages have a surprisingly high reach with some packages have a reach of \textbf{27299} packages, the \textbf{middle quantile} falls at 27. The average is \textbf{471} packages.

\subsubsection{Prevention in the wild}
TypoSquatting attempts are common and most organizations are aware of it. Some companies use it as a defense by registering packages with names Amazon for it's package \textbf{aws-encryption-sdk} registers 15 other packages. They mainly handle hyphenation errors (aws-encryption-sdk) and spelling mistakes (awsencrptyion). Similarly, Amazon typosquats 23 packages for \textbf{aws-encryption-sdk-cli} and 8 for \textbf{dynamodb-encryption-sdk}. A manual verification tells us that no other author in the top 10 holding the most packages, employs these measures. 

\textbf{The next section defines the second part of our project that the paper discusses. We analyze potential software license violations that may have been caused due to transitive dependencies.}

\section{License violations in PyPI}
\label{sec:license}

\begin{figure}
  \centering
    \includegraphics[width=0.5\textwidth]{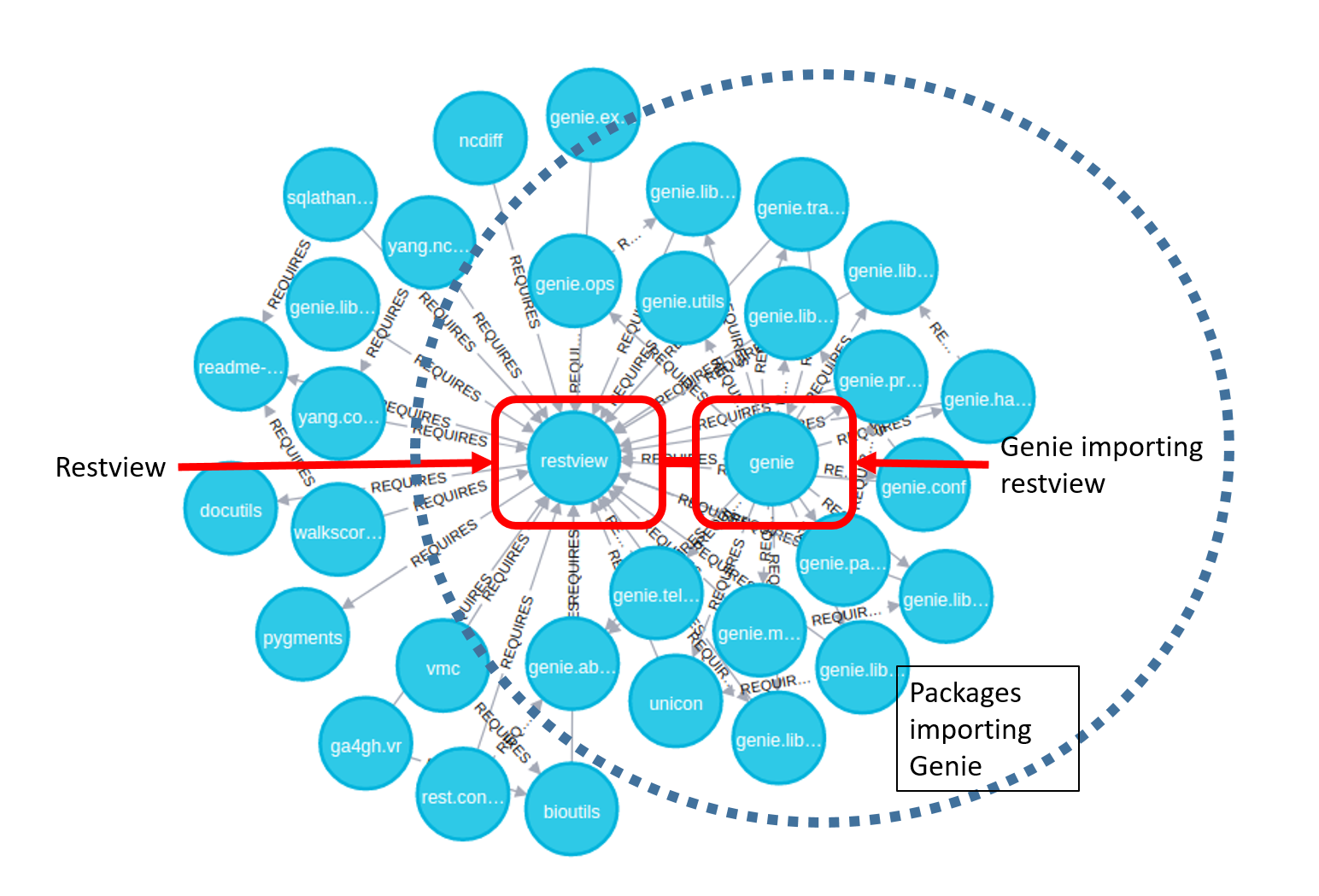}
    \caption{The dependency graph between Cisco's genie and restview. Restview carries a GPL license whereas Genie is released under Apache causing a violation. All packages importing Genie unknowingly cause a violation too.}
    \label{fig:restview_genie}
\end{figure}

\begin{figure}
  \centering
    \includegraphics[width=3.25in]{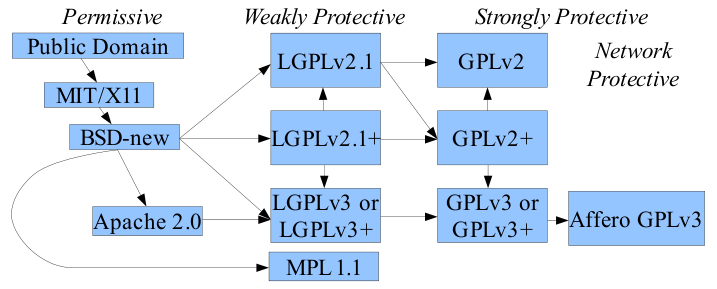}
    \caption{The Open Source License Compatibility Graph, Source: \cite{licensegraph}}
    \label{fig:license_graph}
\end{figure}

We analyzed the PyPI ecosystem for open source licensing violations between dependent packages. Figure \ref{fig:license_graph} shows the compatibility for varying open source licenses. Software with more protective licenses may incorporate other components under less protective licenses, while license violations occur when software with less protective licenses include components protected by stronger licenses. In our ecosystem, we consider that a violation has occurred if a package x imports another package y if package y has a lesser permissible license than package x. PyPI performs no automated checks on OSS license violations. In fact, PyPI allows license to be a free text field which resulted in more than 10,000 unique strings used in the license fields. For our analysis, we combined popular OSS licenses (For example, mit license and MIT) into single entities. OSS License violation are a serious problem since a violation caused by a package can also transitively result in violations in other packages that implicitly trust this package.  

\subsection{Observed Violations}

We identify violations by querying our dependency graph and explicitly looked for packages with permissive licenses that include packages with restrictive licenses. Our findings for certain violation types is shown in Table \ref{tab:license_violations}. The frequent violations from packages with MIT licenses is most likely due to the popularity of MIT licenses within PyPI.

\begin{table}
\centering
    \begin{tabular}{l|l}
    \textbf{Violation Type} &  \textbf{Occurrences} \\ \hline
    MIT importing GPLv3 &    519  \\
    BSD importing GPLv3 &   80  \\
    Apache 2.0 importing GPLv3 &     73  \\\hline
    MIT importing LGPLv3 &   691  \\
    BSD importing LGPLv3 &   92  \\
    Apache 2.0 importing LGPLv3 & 49 \\
    \hline
    \end{tabular}
    \caption{Different types of license violations present on PyPI.}
    \label{tab:license_violations}
\end{table}

One instance of license violation we inspected was Cisco's Genie, a test automation framework mainly for internal use, which is permissively licensed under Apache, but Genie requires the GPLv3 licensed restview, a package for viewing ReStructuredText documents through the browser. 
By the package dependency graph in Figure \ref{fig:restview_genie}, we can further conclude all the packages importing genie also inherits the same license violation it has with restview. 

\begin{figure}
  \centering
    \includegraphics[width=\linewidth]{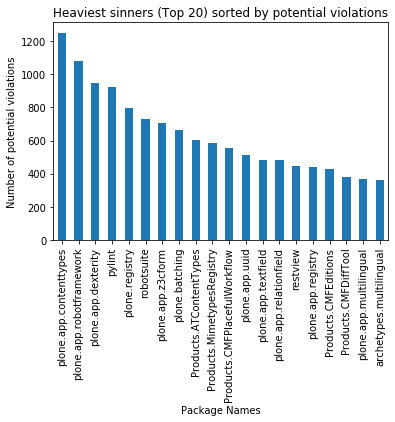}
    \caption{Top 20 packages responsible for most number of potential OSS license violations}
    \label{fig:heaviest_sinners}
\end{figure}

We also enlist packages which are involved in most number of license violations. These packages possess restrictive licenses but have been imported by packages which have been released with permissive licenses. \hyperref[fig:heaviest_sinners]{\ref{fig:heaviest_sinners}} shows the top 20 packages causing the most violations and falling in this category.

In the next section, we discuss certain methods that can be used to protect the ecosystem.

\section{Prevention}
Based on the pattern of package vulnerabilities exposed our study, most issues would be prevented by either one or a combination of these practices. PyPI should make it compulsory to specify dependencies in metadata of uploaded packages. Using this information, PyPI can automate detection of license violations. A permission model similar to mobile phones might be useful when packages are installing. For example, if a package says, ``math.py needs to access your microphone. Grant permission?", the users can instantly spot that something is wrong. Having a trusted maintainer badge / trusted package badge on popular packages might be helpful similar to the one used by popular social networking sites like Twitter, Instagram, etc. It can show statistics about a package when installing. This can help in preventing malicious downloads of popular packages, which are usually most attacked. \pip can alert users of unmaintained / old packages or if a vulnerability has been found in their dependencies. The license field must not be free text. This will make it easier for \pip to automatically identify potential license violations. \cite{pyup} has automated ways of identifying if dependencies carry compatible licenses.

\section{Conclusion and Future Work}
In this project, we presented a large-scale study of the Python ecosystem from the densely connected structure of PyPI. The overall conclusion is that PyPI possesses security risks which need immediate attention. We first analyse the ecosystem to identify the attack vector. 

We show the ease with which the ecosystem can be exploited, analyse the impact of exploiting a package and show methods used by attackers to trick users into downloading malicious packages. We show the number of maintainers/packages affected by an exploit and how many packages/maintainers are implicitly trusted on average while installing a package. We also show the impact of attacking a package or hijacking a maintainers account. 

We also provide examples to quantify the impact of an exploit by taking example of a popular package. We then analyse a real-world exploit to assert that the threat is real. Finally, we also use our data to empirically analyze the problem of open source license violations in the ecosystem.

\bibliographystyle{abbrv}
\bibliography{vldb_sample}  

\end{document}